# BoGEMMS: the Bologna Geant4 multi-mission simulator

A. Bulgarelli*[a], V. Fioretti[b], P. Malaguti[a], M. Trifoglio[a], F. Gianotti[a]

[a] INAF/IASF Bologna, Via Gobetti 101, 40129 Bologna, Italy;
[b] CIFS–Torino, Viale Settimio Severo 3, I-10133, Torino, Italy

*andrea.bulgarelli@inaf.it; phone +39 051 6398774; fax +39 051 6398724; http://www.iasfbo.inaf.it

## ABSTRACT

BoGEMMS, (Bologna Geant4 Multi-Mission Simulator) is a software project for fast simulation of payload on board of scientific satellites for prompt background evaluation that has been developed at the INAF/IASF Bologna. By exploiting the Geant4 set of libraries, BoGEMMS allows to interactively set the geometrical and physical parameters (e.g. physics list, materials and thicknesses), recording the interactions (e.g. energy deposit, position, interacting particle) in *NASA FITS* and *CERN root* format output files and filtering the output as a real observation in space, to finally produce the background detected count rate and spectra. Four different types of output can be produced by the BoGEMMS capturing different aspects of the interactions. The simulator can also run in parallel jobs and store the results in a centralized server via xrootd protocol. The BoGEMMS is a multi-mission tool, generally designed to be applied to any high-energy mission for which the shielding and instruments performances analysis is required.

## 1. INTRODUCTION

Before the advent of particle transport codes, and the increase of the computing performances, the background of a space mission was predicted by means of semi-empirical methods and scaling laws on the basis of the background experienced by the operative missions and ground experiments (see e.g. [1]). This limitations of such approach, due to the different design of the instruments and the complexity of particle interactions, led to big uncertainties in the final background evaluation, well far from the accuracy needed in the reconstruction of the true source signal. The development of Monte Carlo based simulations of the particles interaction with matter, coupled with computer 3D models of the spacecraft and instruments design, allows to track the particles and their secondary particles from the first hit to the final energy deposit on the detection plane, producing the spatial and energy distribution of the detected background counts with the resolution of a real observation in space ([2], [3]). The possibility of creating a virtual model of the telescope and exposing it to the space radiation environment has fundamental benefits along the entire project development: the shielding optimization, the production and characterization of the background flux, the calibration validation, the treatment and filtering of the observation data sets.

The accuracy of the background Monte Carlo simulations depends on many factors:

- the modeling of the space radiation environment;
- the reliability of the interaction cross sections and parameterization of the physics processes;
- the building of the spacecraft and instruments geometry and composition model.

Among many nuclear and particles physics codes, we chose the open source C++ based Geant4 Monte Carlo toolkit ([4], [5]). Developed by CERN and maintained by a large, international collaboration. The Geant4 toolkit was first conceived for the high energy experiments involved at particles accelerators and then extended to "lower" energy ranges, i.e. the X-ray and γ-ray domain, and it is now a widely used particle transport code for astrophysics missions. The code not only provides a set of classes and libraries to describe the interactions of particles with matter down to low energies (250 eV), but it also allows to easily build the 3D model of the spacecraft and payload and to generate the energy and spatial distribution of the particles in space. Validation testing results can be found in [6] and [7].

On the basis of the Geant4 set of libraries, the BoGEMMS, Bologna Geant4 Multi-Mission Simulator, project for prompt background evaluation has been developed at the INAF/IASF Bologna, which allows to interactively set the geometrical and physical parameters (e.g. physics list, materials and thicknesses), recording the interactions (e.g. energy deposit, position, interacting particle) in *NASA FITS* [8] and *CERN root* [9] format output files and filtering the output as a real observation in space, to finally produce the background detected count rate and spectra. The BoGEMMS is a multi-mission tool, generally designed to be applied to any high-energy mission for which the shielding and instruments performances analysis is required.

Until now, this simulator has been used for the detection efficiency and background evaluation of the Simbol-X [10], NHXM [11] and Gamma-Light missions [12], as well as for the XMM-Newton soft protons impact evaluation [13] and the general analysis of the shielding efficiency in LEO [14].

## 2. THE SIMULATOR CONFIGURATION

The Geant4 software provides all the libraries needed to perform the simulation, but the user has the responsibility to choose the physics needed for its virtual experiment and to select which kind of information must be provided (e.g. the energy deposit, the particle energy or direction, the type of interaction), all features that can be modified only in the hard-coded Geant4 working files. As a direct consequence, for each new experiment a new simulator has to be developed.

The BoGEMMS solves this problem by allowing the user to set the simulator by means of a configuration file read by the simulator. The general features of the configuration file are:
- the setting of the physics list (e.g. electromagnetic, hadronic, both);
- the selection of the output file among four types, providing information on the particle energy and track, or the volumes energy deposit, or the physics process in play for each step, and the possibility to write only a set of volumes to speed up the analysis;
- in case of simple geometries, the possibility to build directly the instruments and shielding design, as well as the creation of empty non interacting volumes (called "stepping volumes") to track or remove the particles.

In addition to the configuration file, the simulator takes as input the energy, angular and spatial distribution of the impinging particles, efficiently coded in a single file provided by the Geant4 General Particles Source (GPS) tool [15].

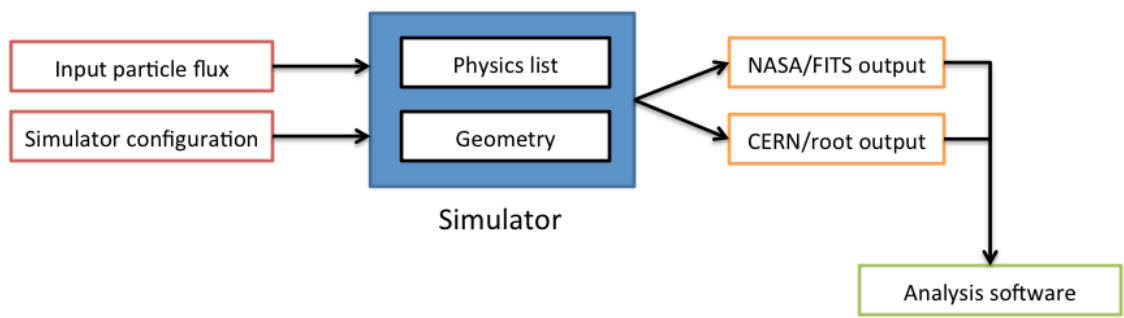

Figure 1. The BoGEMMS architecture: the simulator, composed by the physics and geometry libraries, receives in input the particle flux and the configuration file setting the geometry, the physics processes and the output file, that is finally produced in *NASA FITS* or *CERN root* format and read by an analysis software.

The BoGEMMS architecture, schematized in Figure 1, can be divided in three main steps:
- the input files where the user sets the emitted particles and the simulator physics, geometry and output files (format and content);

- the BoGEMMS core architecture, where the Geant4 libraries are called according to the input files and the outputs produced in *NASA FITS* or *CERN root* format;
- an analysis software, that filters the output and produces, for example, the background and spectral files of various payload components.

In addition, when the *CERN root* output format file is selected, it is possible to send the output of the simulator to a remote machine via xrootd. With this mechanism it is possible to run many simulators in parallel and store the simulation results in a unique data server. The main advantage of this approach is obvious: it can centralize the data storage while maintaining the ability to run many jobs in parallel on different machines of a cluster without having to use a distributed file system.

**2.1 Physics processes**

A detailed list of all the particles and physics processes activated in BoGEMMS is outside the scope of this paper, but we have different physics list optimized for different energy ranges that take into account both electromagnetic (also with the low energy extension) and hadronic processes. In the current version of the simulator the radioactive decay is not taken into account, since we are only evaluating the prompt background level.

The Geant4 releases up to the 9.3 require the user to choose between the standard electromagnetic processes, faster but unable to well reproduce the low energy (E< 100 keV) domain, and the low energy extension, which introduces the atomic relaxation (X- ray fluorescence and Auger electrons emission). The latter is used for the present simulations when our goal is to evaluate the X-ray background.

**2.2 Geant4 output**

The Geant4 output is produced in *NASA FITS* files, a widely used format in the Astrophysical data analysis field, or in *CERN root* format, a widely used format in the High Energy physics community. Four different types of output can be produced by the BoGEMMS:

- IN-OUT type: the particles properties (type, energy, direction, position and interacting volume) are written at the entrance and exit of each touched volume for the entire path. If secondary particles are produced, they are also written starting from the production volume. This type of file is useful to study, for example, the particles interaction with the detection plane and generate the background counts or to trace the properties of the protons after they interact with the X-ray optics.
- ENERGY type: the sum of the energy deposits, due to the sum of primaries and secondary particles, for each volume and each emitted particle is written. This allows to produce smaller files, easier to be handled by the analysis software, used to evaluate the instruments and active shield detected counts;
- STEP type: the Geant4 simulator follows the particles for each interaction step, allowing to write the physics process in play, as well as the particle properties during the interaction. This output type can be used only for a small number of input particles, given the big amount of data, and it is used to test and validate the simulation.
- XYZ type: collects the hits of the activated sensitive volumes. When the particle enters the volume, the energy deposit is summed until the particle exits the volume, or it is absorbed/converted. If a secondary is produced within the sensitive volume, its energy deposit is also recorded along its path. A new feature in the sensitive detector class has been added for the analysis of tracks in Compton/pair production telescopes (as in the case of Gamma-light proposal). When a secondary is produced, if it is generated by a Compton scattering or a gamma conversion a flag is added according to the process. In addition, both primary and secondary photons that Compton scatter in the active tracker layers are flagged in the output file. In the configuration file is possible not only to select which volumes writing in output, but also the tracker layers.

## 3. SOME GEOMETRY EXAMPLES

The Geant4 output is treated as a real observation in space, from the screening phase (e.g. reconstruction of the detected count, anticoincidence triggering) to the differential spectra production in counts cm$^{-2}$ s$^{-1}$ keV$^{-1}$.

For example, in the analysis of the Simbol-X and NHXM background levels (see Figure 2), the energy ranges of the LED (Low Energy Detector) and the HED (High Energy Detector) are respectively set to 0.5 - 20 keV and 5 - 100 keV, while in the XMM-Newton simulation the soft protons induced counts are selected in the 1 - 100 keV energy range.

Two additional features are also possible in the analysis process: the evaluation of the first volume hit by the primary particle, to look for a leakage in the shielding and to select the volumes responsible for the highest background fraction, and recording the energy distribution of the particles finally interacting with the detection plane.

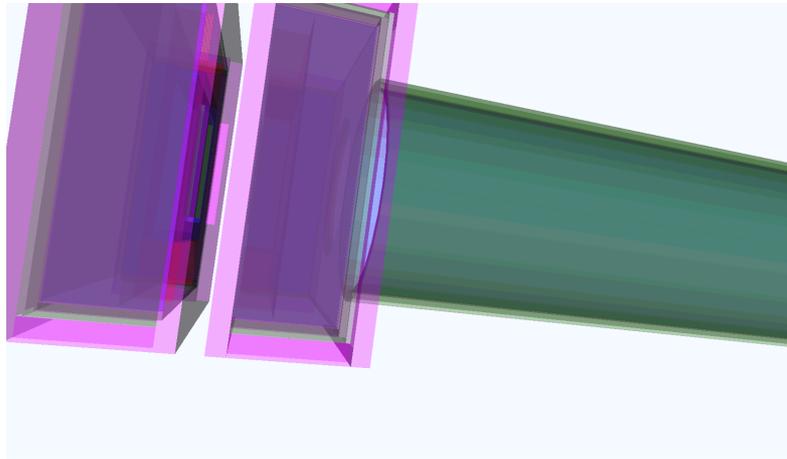

Figure 2. The geometry developed for the NHXM ESA proposal. The LED and HED detectors with shielding systems (on the left) and the collimator (on the right) are shown. No spacecraft structures are shown. The shielding system is compound by BGO (20 mm thickness), polystirene (3 mm thickness) and a graded shield.

Another example is the Gamma-Light mission. Gamma-Light is a Compton/pair production based gamma-ray mission to be proposed to the ESA call for a small mission opportunity for a launch in 2017. Supported by a joint Italian-European high energy Astrophysics community, the telescope goal is to observe the 10 – 100 MeV energy with unprecedented sensitivity and angular resolution. Figure 3 shows the geometry configuration of Gamma-Light.

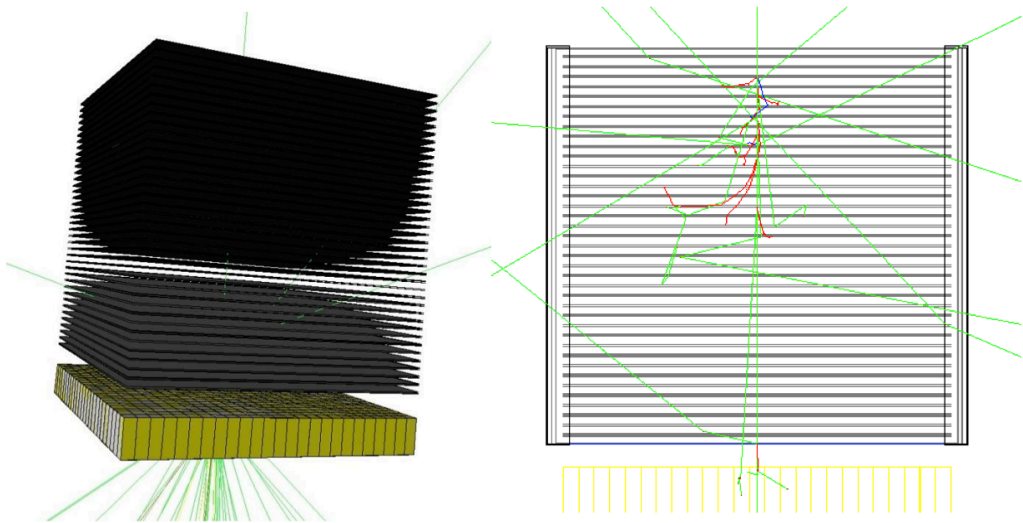

Figure 3. Gamma-Light geometry configuration. Left panel: View of the tracker Si layers (in gray) and, at the bottom, the CsI based calorimeter (in yellow) as built by BoGEMMS. Right panel: Lateral view of the photons (in green), electrons (in red) and positrons (in blue) generated by a 10 MeV monochromatic beam (10 summed primary photons) along the tracker. The lateral bars represent the electronics surrounding the tracker.

## 4. CONCLUSION

The BoGEMMS is a multi-mission tool designed to be applied to any high-energy mission for which the shielding and instruments performances analysis is required. We have tested the simulator with many different configuration of some proposal (Simbol-X, NHXM, Gamma-Light). The software has proven to be flexible and allowed us to build simulations in a very short time. We only needed just to concentrate our effort on the development of geometry and in the study of the background model as input, while all other aspects (events tracking, content and format of output files, management of parallel runs, automatic storing of the results on a centralized data server) were already provided by the software architecture simulator itself. With this architecture it is possible to focus the efforts on the mission and payloads to be simulated (testing different payload configurations) and not in the software development of the simulator itself.